# Limited Diffraction Solutions to
# Maxwell and Schroedinger Equations

Jian-yu Lu[*], Ph.D., J.F.Greenleaf[*], Ph.D., and Erasmo Recami[**], Ph.D.

[*]Biodynamics Research Unit, Department of Physiology and

Biophysics, Mayo Clinic and Foundation, Rochester, MN 55905, U.S.A.

[**]Facolta' di Ingegneria, Universita' statale di Bergamo, Dalmine

(BG), Italy; INFN - Sezione di Milano, Milan, Italy; and D.M.O.,

FEE, State University at Campinas, Campinas, S.P., Brazil.

**Abstract** — We have developed a new family of limited diffraction *electromagnetic* X-shaped waves based on the scalar X-shaped waves discovered previously. These waves are diffraction-free in theory and particle-like (wave packets), in that they maintain their shape as they propagate to an infinite distance. The "X waves" possess (theoretically) infinitely extended "arms" and —at least, the ones studied in this paper— have an infinite total energy: therefore, they are not physically realizable. However, they can be truncated in both space and time and "approximated" by means of a finite aperture radiator so to get a large enough depth of interest (depth of field). In addition to the Maxwell equations, X wave solutions to the free Schroedinger equation are also obtained. Possible applications of these new waves are discussed. Finally, we discuss the appearance of the X-shaped solutions from the purely geometric point of view of the special relativity theory.

# I. — INTRODUCTION

"Limited diffraction" beams were discovered as early as 1941 by Stratton[1] and rediscovered by Durnin in 1987 [2,3]. These beams have an infinite depth of field, i.e., they can propagate to an infinite distance without changing their wave shape. Durnin termed these beams "nondiffracting beams"[2] or "diffraction-free beams"[3]. Because Durnin's terminology is controversial, we used the term "limited diffraction beams" based on the fact that in practice all beams will diffract eventually[4,5]. Durnin's beams are also called Bessel beams because their transverse beam profile is a Bessel function[2]. Bessel beams are monochromatic and are obtained by treating the (scalar) amplitude related to one transverse component only of either the electric or the magnetic field of light as a solution to the scalar wave equation[2]. Theoretically, the limited diffraction beams considered in this paper have an infinite total energy; moreover they would require an infinite aperture radiator to be produced. In reality, however, these beams can be truncated in both space and time and can be "approximated" by means of a *finite aperture* radiator so to get a large enough depth of interest (a large field depth)[3]. Because of this property, limited diffraction beams could have applications in medical imaging[4–14], tissue characterization[15], Doppler velocity estimation[16], non-destructive evaluation of materials[17], and other related physics areas such as electromagnetism[18–20] and optics[2,3,21], besides —possibly— geophysics (seismic waves) and even gravitational wave detection (see the following).

Because limited diffraction beams have many potential applications, they have been studied extensively in recent years in both acoustics[22–35] and optics[2,3,21]. Recently, we have discovered a new family of limited diffraction beams[36]. These beams have been called "X waves" because they are X-shaped in a plane passing through the propagation axis ($r - z$ plane)[37–39]. The X-shaped waves are different from the Bessel beams[36] because they contain multiple frequencies and are nondispersive in isotropic-homogeneous media or free space (Bessel beams are "limited diffraction" at a single frequency, but become dispersive for multiple frequencies because the phase velocity of each frequency component is different[9]); moreover, the X-shaped waves are Superluminal[36,37,40—42], i.e., have Superluminal group velocities, while Bessel beams are subluminal as it will be shown elsewhere.

In cylindrical coordinates, limited diffraction beams propagating along the $z$ axis can be written in the following form

$$\Phi(r, \phi, z - c_1 t),\tag{1}$$

where $r, \phi, z$, and $t$ represent the radial distance, polar angle, axial distance, and time, respectively, $\Phi$ represents acoustic pressure, velocity potential, or Hertz potential, $z - c_1 t$ is a propagation term, and $c_1$ is the velocity of the beam. Because the variables, $z$ and $t$, appear only in the propagation term in eq.(1), limited diffraction beams are only a function of $r$ and $\phi$ if $z - c_1 t =$ constant, i.e., traveling with the beam at the speed of $c_1$, one sees a constant beam pattern. This is different from conventional focused beams[43] and localized waves studied by Brittingham[18] and other investigators[19–20,44].

In this paper, we will extend the theory of limited diffraction beams to electromagnetic waves. X-shaped wave solutions to the free Maxwell and the free Schroedinger equations will be obtained.

## II. — LIMITED DIFFRACTION SOLUTIONS TO FREE MAXWELL EQUATIONS

### A. Maxwell's Wave Equations

In this Section we will present exact limited diffraction solutions to the free-space Maxwell equations[45]:

$$\nabla \times \boldsymbol{H} = \varepsilon_0 \frac{\partial \boldsymbol{E}}{\partial t}, \qquad (2)$$

$$\nabla \times \boldsymbol{E} = -\mu_0 \frac{\partial \boldsymbol{H}}{\partial t}, \qquad (3)$$

$$\nabla \cdot \boldsymbol{E} = 0, \qquad (4)$$

and

$$\nabla \cdot \boldsymbol{H} = 0, \qquad (5)$$

where $\boldsymbol{E}$ is the electric field strength, $\boldsymbol{H}$ is the magnetic field strength, $\epsilon_0$ is the dielectric constant of free space ($\epsilon_0 \approx \frac{1}{36\pi} \times 10^{-9}$ F/m), $\mu_0$ is the magnetic permeability of free space ($\mu_0 = 4\pi \times 10^{-7}$ H/m), and $t$ is time.

From eqs.(2) to (5), one obtains the free wave equations

$$\nabla^2 \boldsymbol{E} - \frac{1}{c^2} \frac{\partial^2 \boldsymbol{E}}{\partial t^2} = 0 \qquad (6)$$

or

$$\nabla^2 \boldsymbol{H} - \frac{1}{c^2} \frac{\partial^2 \boldsymbol{H}}{\partial t^2} = 0, \qquad (7)$$

where $\nabla^2$ represents the three-dimensional Laplacian operator, and $c = 1/\sqrt{\varepsilon_0 \mu_0}$ is the speed of light in free space ($c \approx 3 \times 10^8$ m/s). Note that $\boldsymbol{E}$ and $\boldsymbol{H}$ in eqs.(6) and (7) are related by eqs.(2) and (3).

### B. Scalar Approximation of Maxwell Wave Equations

Equations (6) and (7) can be written by

$$\begin{cases} \nabla^2 E_x - \frac{1}{c^2} \frac{\partial^2 E_x}{\partial t^2} = 0 \\ \nabla^2 E_y - \frac{1}{c^2} \frac{\partial^2 E_y}{\partial t^2} = 0 \\ \nabla^2 E_z - \frac{1}{c^2} \frac{\partial^2 E_z}{\partial t^2} = 0 \end{cases} \qquad (8)$$

and

$$\begin{cases} \nabla^2 H_x - \frac{1}{c^2}\frac{\partial^2 H_x}{\partial t^2} = 0 \\ \nabla^2 H_y - \frac{1}{c^2}\frac{\partial^2 H_y}{\partial t^2} = 0 \\ \nabla^2 H_z - \frac{1}{c^2}\frac{\partial^2 H_z}{\partial t^2} = 0 \end{cases}, \qquad (9)$$

respectively, where each component of $\boldsymbol{E}$ and $\boldsymbol{H}$ is coupled with the others through the Maxwell equations (2)-(5). Because of the coupling of the components, it is difficult to solve eqs.(8) and (9) directly. However, in some cases such as optics and microwaves[46], the equations can be simplified, i.e., only the scalar amplitude of one transverse component of either $\boldsymbol{E}$ or $\boldsymbol{H}$ is considered and any other components of interest are treated independently in a similar fashion (treating light and microwaves as a scalar phenomenon). This is approximately true under the following conditions[46]: (i) the diffracting aperture must be large compared with a wavelength, and (ii) the diffracted fields must not be observed too close to the aperture. In this case, limited diffraction beams developed in acoustics[4–20] can be directly applied to electromagnetism because they share the same wave equation. This was verified experimentally in optics by Durnin for Bessel beams[3].

*C. Hertz Vector Potential*

Another way to solve the Maxwell equations is to use a Hertz vector potential. This approach is *rigorous* as opposed to the scalar method above.

Because of eq.(4), the electric field strength can be written[20]

$$\boldsymbol{E} = -\mu_0 \frac{\partial}{\partial t}\nabla \times \boldsymbol{\Xi}_m, \qquad (10)$$

where $\boldsymbol{\Xi}_m = \Phi \boldsymbol{n}^o$, is the magnetic Hertz vector potential where $\boldsymbol{n}^o$ represents a unit vector. This implies that the electromagnetic wave given by eq.(4) is a TE polarization wave that is perpendicular to $\boldsymbol{n}^o$ (for TM polarization, the analysis is similar). From

eq.(3), we obtain

$$\boldsymbol{H} = \nabla \times (\nabla \times \boldsymbol{\Xi}_m).$$ (11)

By inserting eq.(10) into eq.(2), and using eq.(11) and the Lorentz condition, $\nabla \cdot \boldsymbol{\Xi}_m = -f_m$ [$\boldsymbol{\Xi}_m$ is not unique: i.e., different $\boldsymbol{\Xi}_m$ may all give the same $\boldsymbol{E}$ and $\boldsymbol{H}$ from eqs.(10) and (11), respectively], where $f_m$ is any differentiable scalar function, one obtains the wave equation that the magnetic Hertz vector potential must satisfy:

$$\nabla^2 \boldsymbol{\Xi}_m - \frac{1}{c^2}\frac{\partial^2 \boldsymbol{\Xi}_m}{\partial t^2} = 0.$$ (12)

From eqs.(10), (11) and (12), we obtain $\boldsymbol{E}$ and $\boldsymbol{H}$ that solve eqs.(6) and (7), respectively.

If we use cylindrical coordinates and let $\boldsymbol{n}^o = \boldsymbol{z}^o$, where $\boldsymbol{z}^o$ is a unit vector along the $z$ axis, from eqs.(10) and (11) we get

$$\boldsymbol{E} = -\mu_0 \frac{1}{r}\frac{\partial^2 \Phi}{\partial t \partial \phi}\boldsymbol{r}^o + \mu_0 \frac{\partial^2 \Phi}{\partial t \partial r}\boldsymbol{\phi}^o$$ (13)

and

$$\boldsymbol{H} = \frac{\partial^2 \Phi}{\partial r \partial z}\boldsymbol{r}^o + \frac{1}{r}\frac{\partial^2 \Phi}{\partial \phi \partial z}\boldsymbol{\phi}^o + \left(\frac{\partial^2 \Phi}{\partial z^2} - \frac{1}{c^2}\frac{\partial^2 \Phi}{\partial t^2}\right)\boldsymbol{z}^o,$$ (14)

respectively, where $\Phi$ is a solution to the free scalar wave equation (obtained directly from eq.(12)),

$$\left(\nabla^2 - \frac{1}{c^2}\frac{\partial^2}{\partial t^2}\right)\Phi = 0,$$ (15)

and where $\boldsymbol{r}^o$ and $\boldsymbol{\phi}^o$ are the unit vectors along the variables, $r$ and $\phi$, respectively. From eq.(15), eqs.(6) and (7) can be solved.

*D. Limited Diffraction Solutions to Maxwell Equations*

Equation (6) or (7) can either be solved approximately by treating each component of the vectors independently, or by using the Hertz vector potential. However, the solutions are not necessarily limited diffraction. In the following, we will show how to obtain limited diffraction solutions.

From eqs.(10) and (11), we see that $\boldsymbol{E}$ and $\boldsymbol{H}$ are related to $\Phi$ by derivatives of its variables, $r, \phi, z,$ and t, respectively. This means that if $\Phi$ is a limited diffraction solution to eq.(15), the solution to the Maxwell wave equations is also limited diffraction. This is because the derivatives with respect to the variables do not change the propagation term, $z - c_1 t$.[4] If the scalar method is used, it is straightforward to obtain limited diffraction solutions because each scalar component of $\boldsymbol{E}$ or $\boldsymbol{H}$ satisfies the same equation as $\Phi$ (eqs.(8) and (9)). Because of this, numerous limited diffraction (relativistic) electromagnetic waves can be obtained from the scalar limited diffraction (non-relativistic) beams studied in acoustics[4].

Families of generalized solutions of eq.(15) were discovered recently[36]. One of the families of solutions is given by[36]:

$$\Phi_\zeta(s) = \int\limits_0^\infty T(k) \left[ \frac{1}{2\pi} \int\limits_{-\pi}^\pi A(\theta) f(s) d\theta \right] dk, \tag{16}$$

where

$$s = \alpha_0(k, \zeta) r \cos(\phi - \theta) + b(k, \zeta)[z \pm c_1(k, \zeta) t], \tag{17}$$

and where

$$c_1(k, \zeta) = c \sqrt{1 + [\alpha_0(k, \zeta)/b(k, \zeta)]^2}. \tag{18}$$

$T(k)$ is any complex function (well behaved) of $k$ and could include the temporal frequency transfer function of a practical acoustic transducer or electromagnetic antenna, $A(\theta)$ is any complex function (well behaved) of $\theta$ and represents a weighting function of the integration with respect to $\theta$, quantity $f(s)$ is any complex function (well behaved) of $s$, $\alpha_0(k, \zeta)$ and $b(k, \zeta)$ are any complex function of $k$ and $\zeta$, $c$ is the speed of sound or light in eq.(15), $k$ and $\zeta$ are variables that are independent of the spatial position, $\boldsymbol{r} = (r\cos\phi, \ r\sin\phi, \ z)$, and time, $t$, and $\zeta$ is an Axicon angle $(0 < \zeta < \pi/2)$[36].

If $c_1(k, \zeta)$ in eq.(18) is real, "$\pm$" in eq.(17) represent forward and backward propagating waves, respectively (in the following analysis, we consider only the forward propagating waves and all results will be the same for the backward propagating waves). Furthermore, $\Phi_\zeta(s)$ will represent a family of limited diffraction waves if $c_1(k, \zeta)$ is independent of $k$ (containing the same propagation terms, $z - c_1(\zeta)t$, for all frequency components, $k$).

It must be noticed that $\Phi_\zeta(s)$ in eq.(16) is very general. It contains some of the limited diffraction solutions known previously, such as: the plane wave, Durnin's limited diffraction beams and the limited diffraction portion of the Axicon beam, in addition to an infinity of new beams[36].

## III. — ELECTROMAGNETIC X-SHAPED WAVES

In this Section, we will derive novel limited diffraction electromagnetic X-shaped waves from eq.(16) by the Hertz vector potential, $\boldsymbol{\Xi}_m = \Phi \boldsymbol{z}^{\,o}$, and by eqs.(13) and (14).

### A. X-Shaped Wave Solutions

Let $T(k) = B(k)e^{-a_0 k}$, $A(\theta) = i^n e^{in\theta}$, $\alpha_0(k, \zeta) = -ik\sin\zeta$, $b(k, \zeta) = ik\cos\zeta$, and $f(s) = e^s$ in eq.(16); we obtain an nth-order scalar limited diffraction X wave that has an "X-like" shape in a plane passing through the propagation axis ($r - z$ plane)[36],

$$\Phi_{X_n}(r, \phi, z - c_1 t)$$

$$= e^{in\phi} \int_0^\infty B(k) J_n(kr \sin \zeta) e^{-k[a_0 - i \cos \zeta(z - c_1 t)]} dk, \quad (n = 0,\ 1,\ 2,\ ...), \qquad (19)$$

where $B(k)$ is any function (well behaved) of $k$ and represents a transfer function of an acoustic transducer or electromagnetic antenna, $J_n(\cdot)$ is the nth-order Bessel function of the first kind, $c_1 = c/\cos \zeta$, $k = \omega/c$, $\omega$ is the angular frequency, $a_0 > 0$ and $0 \leq \zeta < \pi/2$ are constants.

If $B(k) = a_0$, from eq.(19) we have the nth-order broadband X wave[36],

$$\Phi_{XBB_n}(r, \phi, z - c_1 t) = \frac{a_0(r \sin \zeta)^n e^{in\phi}}{\sqrt{M}\left(\tau + \sqrt{M}\right)^n}, \quad (n = 0,\ 1,\ 2,\ ...), \qquad (20)$$

where the subscript "BB" means "broadband," and $M = (r \sin \zeta)^2 + \tau^2$, and where $\tau = a_0 - i \cos \zeta(z - c_1 t)$.

If $B(k)$ is a band-limited function, we obtain an nth-order band-limited X wave that is a convolution of functions $\mathcal{F}^{-1}\left[B\left(\frac{\omega}{c}\right)\right]/a_0$ and $\Phi_{XBB_n}(r, \phi, z - c_1 t)$ with respect to time, $t$: [36]

$$\Phi_{XBL_n}(r, \phi, z - c_1 t) = \frac{1}{a_0}\mathcal{F}^{-1}\left[B\left(\frac{\omega}{c}\right)\right] * \Phi_{XBB_n}, \quad (n = 0,\ 1,\ 2,\ ...), \qquad (21)$$

where $\mathcal{F}^{-1}$ represents the inverse Fourier transform, $*$ denotes the convolution, and subscript "BL" means "band-limited".

Substituting eq.(20) into eqs.(13) and (14), one obtains an nth-order broadband limited diffraction electromagnetic X wave:

$$(\boldsymbol{E}_{XBB_n})_r = -\frac{n\mu_0 c}{rM}\left(\tau + n\sqrt{M}\right)\Phi_{XBB_n}, \qquad (22)$$

$$(\boldsymbol{E}_{XBB_n})_\phi = \frac{i\mu_0 c}{rM}\Phi_{XBB_n}$$

$$\cdot \left\{\frac{3\tau + 2n\sqrt{M}}{M}r^2 \sin^2 \zeta + n\left(\tau + n\sqrt{M}\right)\left[\frac{r^2 \sin^2 \zeta}{\sqrt{M}\left(\tau + \sqrt{M}\right)} - 1\right]\right\}, \qquad (23)$$

$$(\boldsymbol{H}_{XBB_n})_r = -\frac{\cos\zeta}{\mu_0 c}(\boldsymbol{E}_{XBB_n})_\phi, \tag{24}$$

$$(\boldsymbol{H}_{XBB_n})_\phi = \frac{\cos\zeta}{\mu_0 c}(\boldsymbol{E}_{XBB_n})_r, \tag{25}$$

$$(\boldsymbol{H}_{XBB_n})_z = \frac{\sin^2\zeta}{M^2}\left[\left(n^2-1\right)M + 3\tau\left(\tau + n\sqrt{M}\right)\right]\Phi_{XBB_n}, \tag{26}$$

$$(n = 0,\ 1,\ 2,\ \cdots).$$

Because Maxwell equations are linear, both real and imaginary parts of their solutions are also solutions. In the following, we consider the real part only.

### B. Poynting Flux and Energy Density

From eqs.(22)-(26), one obtains the Poynting flux, $\boldsymbol{P}_{XBB_n} = Re\{\boldsymbol{E}_{XBB_n}\} \times Re\{\boldsymbol{H}_{XBB_n}\}$, and the energy density, $U_{XBB_n} = \epsilon_0|Re\{\boldsymbol{E}_{XBB_n}\}|^2 + \mu_0|Re\{\boldsymbol{H}_{XBB_n}\}|^2$, of the nth-order limited diffraction electromagnetic X waves:

$$(\boldsymbol{P}_{XBB_n})_r = Re\left\{(\boldsymbol{E}_{XBB_n})_\phi\right\}Re\{(\boldsymbol{H}_{XBB_n})_z\}, \tag{27}$$

$$(\boldsymbol{P}_{XBB_n})_\phi = -Re\{(\boldsymbol{E}_{XBB_n})_r\}Re\{(\boldsymbol{H}_{XBB_n})_z\}, \tag{28}$$

$$(\boldsymbol{P}_{XBB_n})_z = \frac{\cos\zeta}{\mu_0 c}\left[|Re\{(\boldsymbol{E}_{XBB_n})_r\}|^2 + \left|Re\left\{(\boldsymbol{E}_{XBB_n})_\phi\right\}\right|^2\right], \tag{29}$$

and

$$U_{XBB_n} = \epsilon_0\left(1 + \cos^2\zeta\right)\left[|Re\{(\boldsymbol{E}_{XBB_n})_r\}|^2 + \left|Re\left\{(\boldsymbol{E}_{XBB_n})_\phi\right\}\right|^2\right]$$
$$+ \mu_0|Re\{(\boldsymbol{H}_{XBB_n})_z\}|^2,\ (n = 0,\ 1,\ 2,\ ...). \tag{30}$$

The total energy of the nth-order limited diffraction electromagnetic X waves is given by

$$U_{XBB_n}^{tot} = \int\limits_{-\pi}^{\pi} d\phi \int\limits_{-\infty}^{\infty} dz \int\limits_{0}^{\infty} r dr\, U_{XBB_n}, \tag{31}$$
$$(n = 0,\ 1,\ 2,\ ...),$$

which is infinite because the decay of the energy density along the X branches[4,36] approaches $1/|z - \frac{c}{\cos\zeta}t|$. Nevertheless, limited diffraction X waves have a finite energy

density and can be approximately produced with a finite aperture and energy over large distance of interest (large depth of field)[37].

## C. An Example

In the following, we give an example of electromagnetic X-shaped wave. For simplicity, only the zero[th]-order ($n = 0$) X wave is considered. Notice that for $n = 0$, eqs.(20)-(26) are axially symmetric (not a function of $\phi$), and $(\boldsymbol{E}_{\text{XBB}_o})_r$, $(\boldsymbol{H}_{\text{XBB}_o})_\phi$, and $(\boldsymbol{P}_{\text{XBB}_o})_\phi$ are zero.

The real part of the zero[th]-order scalar Hertz potential (eq.(20)) and the electromagnetic X waves (eqs.(23),(24) and (26)) are shown in Fig.1. The Poynting flux and energy density are shown in Fig.2. Their maxima and minima are summarized in Tables I and II, respectively. From Table I, we see that, with the parameters given in Fig.1, the electromagnetic X waves are almost transverse waves where their axial components are much smaller than those of the transverse components. This is also shown in Table II where the axial component of the Poynting flux is at least four orders larger than its lateral components. Lateral line plots of Figs.1 and 2 along X branches are shown in Fig.3.

## IV. — FINITE APERTURE APPROXIMATION OF X-SHAPED WAVES AND DEPTH OF FIELD

Limited diffraction electromagnetic X waves obtained from eqs.(10) and (11) are exact solutions to the free-space Maxwell wave equations. In these equations, there are no boundary conditions and thus the apertures required to produce the waves are infinite; in addition, the waves have an infinite total energy (eq.(31)); therefore, they cannot be realized with physical devices. However, these waves can be approximated very well over a large depth of field by truncating them in both space and time[36,37]. It is important to notice that the peaks of the truncated X waves move (Superluminally and) almost rigidly along their motion direction over a large depth of interest, without any further support from the antenna.

Because $|\boldsymbol{E}_{X\mathrm{BB}_n}(r, \phi, z - c_1t)| << |\boldsymbol{E}_{X\mathrm{BB}_n}(r, \phi)|$ and $|\boldsymbol{H}_{X\mathrm{BB}_n}(r, \phi, z - c_1t)| << |\boldsymbol{H}_{X\mathrm{BB}_n}(r, \phi)|$ for $|z - c_1t| > d_z/2$ within a finite transverse aperture, where $d_z$ is a constant, X waves can be truncated within an axially moving "window", $[c_1t - d_z/2, \ c_1t + d_z/2]$. The truncated waves do not meet the problems of the theoretical (infinitely extended) ones; for instance, at the aperture surface ($z = 0$) one can always set a new time frame $t' = t + (d_z/2)/c_1$ that starts from $t' = 0$.

If the diameter of the aperture is $D$, the depth of field (defined as the maximum axial distance within which the decrease of the wave magnitude is less than 6-dB with respect to their peak at the aperture surface) of the X wave Hertz potential (eqs.(20) and (21)) is given by[4,6,26,36]

$$Z_{\max} = \frac{D}{2} \frac{1}{\sqrt{\left(\frac{c_1}{c}\right)^2 - 1}} = \frac{D}{2} \cot \zeta. \tag{32}$$

Because the derivatives in eqs.(10) and (11) do not change the Axicon angle, $\zeta$, the depth of field of the electromagnetic X waves produced by the same aperture are the same as that of the Hertz potential. In addition, eq.(32) is also valid for band-limited electromagnetic X waves[4,31]. As an example, if the diameter of the aperture is 20 m and the Axicon angle, $\zeta$, is 0.005°, the depth of field of both broadband and band-limited electromagnetic X waves is 115 km. Simulation[36] of a finite aperture X wave Hertz potential with the Rayleigh-Sommerfeld diffraction formula[46] and its production[37] with an acoustic transducer have been reported in our previous papers.

## V. — X WAVE SOLUTIONS TO THE FREE SCHROEDINGER EQUATION

In addition to Maxwell equations, there are limited diffraction solutions also —for instance— to the Schroedinger wave equation. These solutions are localized and particle like. They may very well be related to the wave nature of quantum particles (cf. also [47]).

The general non-relativistic, time-dependent, and three-dimensional Schroedinger wave equation is given by [48]

$$-\frac{\hbar^2}{2m}\nabla^2\Phi + V\Phi = i\hbar\frac{\partial\Phi}{\partial t}, \tag{33}$$

where $\hbar = h/2\pi$, $h$ is the Planck constant, $m$ is the mass of the particle, $\Phi = \Phi(r, \phi, z, t)$ is a wave function, and $V = V(r, \phi, z, t)$ is the potential. For free objects, when $V = 0$, we have

$$-\frac{\hbar^2}{2m}\nabla^2\Phi = i\hbar\frac{\partial\Phi}{\partial t}. \tag{34}$$

It is easy to prove that if $f(s) = e^s$, eq.(16) is an exact solution to eq.(34), where

$$s = a_0(k, \zeta)r\cos(\phi - \theta) + b(k, \zeta)[z - c_1(k, \zeta)t], \tag{35}$$

and where

$$c_1(k, \zeta) = \frac{\hbar[\alpha_0^2(k, \zeta) + b^2(k, \zeta)]}{i2mb(k, \zeta)}. \tag{36}$$

$\alpha_0(k, \zeta)$ and $b(k, \zeta)$ in eq.(36) are any functions of the free parameters $k$ and $\zeta$, where we assume that $k = 2\pi/\lambda$ is wave number and $\lambda$ is wavelength. Let

$$\alpha_0(k, \zeta) = -ik\sin\zeta \tag{37}$$

and

$$b(k, \zeta) = ik \cos \zeta, \tag{38}$$

from eq.(36) we obtain

$$c_1(k, \zeta) = \frac{\hbar k}{2m \cos \zeta}. \tag{39}$$

From the de Broglie assumption[48], $\hbar k = h/\lambda = p$ where $p = mv$ is the momentum and $v$ is the speed of particle, eq.(39) becomes

$$c_1(k, \zeta) = \frac{v}{2 \cos \zeta}. \tag{40}$$

Substituting eqs.(37), (38) and (40) into eq.(35), and using $f(s) = e^s$, $T(k) = B(k)e^{-a_0 k}$, and $A(\theta) = i^n e^{in\theta}$ in eq.(16), we obtain an nth-order limited diffraction X wave solution to the Schroedinger wave equation (34)

$$\Phi^s_{X_n} = e^{in\phi} \int\limits_0^\infty B(k) J_n(kr \sin \zeta) e^{-k\left[a_0 - i \cos \zeta \left(z - \frac{v}{2 \cos \zeta} t\right)\right]} dk, \tag{41}$$
$$(n = 0, \ 1, \ 2, \ ...),$$

where the superscript "$s$" means "Schroedinger". Similarly, eqs.(20) and (21) are exact nth-order broadband and band-limited limited diffraction X wave solutions also to the Schroedinger wave equation, respectively, after replacing $c$ with $v/2$:

$$\Phi^s_{XBB_n} = \frac{a_0 (r \sin \zeta)^n e^{in\phi}}{\sqrt{M^s}\left(\tau^s + \sqrt{M^s}\right)^n} \ , \quad (n = 0, \ 1, \ 2, ...), \tag{42}$$

and

$$\Phi^s_{XBL_n} = \frac{1}{a_0} \mathcal{F}^{-1}\left[B\left(\frac{\omega}{c}\right)\right] * \Phi^s_{XBB_n}, \ (n = 0, \ 1, \ 2, ...), \tag{43}$$

where $M^s = (r \sin \zeta)^2 + (\tau^s)^2$ and $\tau^s = a_0 - i \cos \zeta \left(z - \frac{v}{2 \cos \zeta} t\right)$. This means that $\Phi^s_{X_n}$ is a new wave function for a free object. In a confined space (as in the case of a free particle passing through a hole of finite aperture), the wave function $\Phi^s_{X_n}$ will change (spread or diffract) beyond a certain distance after the hole.

If $\zeta = 0$, $a_0 = 0$, and $B(k') = \delta(k' - k)$, where $\delta(\cdot)$ is a delta-function, eq.(41) represents a plane wave propagating in the $z$ direction

$$\Phi^s_P = e^{i\left(kz - k\frac{v}{2}t\right)}. \tag{44}$$

Since we are here interested in nonrelativistic free particles, their energy, $E = \hbar\omega$, is the same as the kinetic energy[48], i.e.,

$$E = \frac{1}{2}mv^2. \tag{45}$$

With eq.(45) and the momentum, $p = \hbar k$, we have

$$k\frac{v}{2} = \frac{E}{\hbar}. \tag{46}$$

Substituting eq.(46) into eq.(44), we have the conventional plane wave expression for nonrelativistic free particles traveling in the $z$ direction[48]

$$\Phi^s_P = e^{ikz}e^{-i\frac{E}{\hbar}t} = e^{i(kz - \omega t)}. \tag{47}$$

Because $c_1 = \frac{v}{2\cos\zeta}$, the X waves in eq.(41) could either be slower ($\cos\zeta > 1/2$) or faster ($\cos\zeta < 1/2$) than the particle speed, $v$. For a plane wave ($\zeta = 0$), $c_1 = \frac{v}{2}$ (see eq.(44)). More interesting would be the relativistic case (of Klein-Gordon and Dirac equations).

It might be surprising that solutions of a relativistic, classical equation can be solutions also of a nonrelativistic, quantum equation. However, it is a well-known fact that in the time-independent case the Helmholtz equation and the Schroedinger equation are formally identical[49] (one important consequence of it being that evanescent wave transmission simulates electron tunnelling). In the time-dependent case, such equations become actually different, but nevertheless strict relations still hold between some solutions of theirs, as it will be explicitly shown elsewhere.

## VI. — DISCUSSION

### A. - X Wave Hertz Potential

The scalar X-shaped waves, $\Phi_{XBB_n}$, in eq.(20) satisfies the wave equation (15). It is the component of a Hertz vector potential in the $z$ direction. If $n = 0$, quantity $\Phi_{XBB_n}$ is axially symmetric and has a single peak at the wave center. For $n > 0$, it is zero on the $z$ axis and is axially asymmetric. When both $n$ and $\zeta$ are zero, $\Phi_{XBB_n}$ represents a plane wave. In this case, $\boldsymbol{E}$ and $\boldsymbol{H}$ derived from eqs.(10) and (11) should be zero. The Hertz potential, $\Phi_{XBB_n}$, which may or may not have a physical meaning, is used as an auxiliary function from which new electromagnetic X waves (eqs.(13) and (14)) are derived.

If $\Phi_{XBB_n}$ is treated approximately as one component of $\boldsymbol{E}$ or of $\boldsymbol{H}$ in eqs.(6) or (7), respectively, it has a physical meaning. Many microwave and optical phenomena are treated this way under suitable conditions[46]. A $J_0$ Bessel beam (a special case of X waves[36]) that was treated as one component of the electric field strength of light was produced by Durnin in an optical experiment[3]. In acoustics, $\Phi_{XBB_n}$ represents acoustic pressure or velocity potential[50,51], and has been approximated by a finite aperture acoustic transducer[37] over a large depth of field.

### B. - Nth-order Electromagnetic X Wave

From the Hertz potential, $\Xi_m = \Phi z^{\,o}$, and eqs.(10) and (11), a family of electromagnetic X waves can be obtained (eqs.(22)-(26)). The nonnegative integer $n$ represents in these equations the order of the waves. Because the variable $\phi$ appears in $\Phi_{XBB_n}$ only, $\boldsymbol{E}$ and $\boldsymbol{H}$ have the same axial symmetry as $\Phi_{XBB_n}$. However, for $n = 1$, $\boldsymbol{E}$ and $\boldsymbol{H}$ are not zero on the axis, $z$, and they are not axially symmetric. This means they are not single-valued on the $z$ axis and thus cannot be approximated with a physical device.

For some components of the electromagnetic X waves the field on the axial axis, $z$, is zero. In such a case, there are multiple peaks around the X wave center, and the energy density is high on these peaks. For example, the energy density of the zeroth-order electromagnetic X wave has four sharp peaks (see Panel (3) of Fig.2). This is similar to the case of Ziolkowski's localized electromagnetic wave[20]. Notice that any higher-order ($n > 0$) scalar X-shaped waves are zero on the $z$ axis and also produce multiple peaks[36].

### C. - Other Hertz Vector Potentials

If we choose

$$\Xi_m = \Phi \phi^{\,o},$$  (48)

eqs.(13) and (14) are replaced by

$$\boldsymbol{E} = \mu_0 \frac{\partial^2 \Phi}{\partial t \partial z} \boldsymbol{r}^{\,o} - \mu_0 \left( \frac{1}{r} \frac{\partial \Phi}{\partial t} + \frac{\partial^2 \Phi}{\partial r \partial t} \right) \boldsymbol{z}^{\,o}$$  (49)

and

$$\boldsymbol{H} = \frac{1}{r} \left( \frac{1}{r} \frac{\partial \Phi}{\partial \phi} + \frac{\partial^2 \Phi}{\partial r \partial \phi} \right) \boldsymbol{r}^{\,o} + \left[ \frac{1}{r^2} \left( \frac{\partial^2 \Phi}{\partial \phi^2} + \Phi \right) - \frac{1}{c^2} \frac{\partial^2 \Phi}{\partial t^2} \right] \phi^{\,o} + \frac{1}{r} \frac{\partial^2 \Phi}{\partial z \partial \phi} \boldsymbol{z}^{\,o},$$  (50)

respectively. Substituting the X wave solutions (eq.(20)) into eqs.(49) and (50), we obtain limited diffraction electromagnetic X waves whose electric components stay in a radial plane ($r - z$ plane).

Similarly, if

$$\Xi_m = \Phi \boldsymbol{r}^{\,o},\tag{51}$$

$\boldsymbol{E}$ and $\boldsymbol{H}$ are given by

$$\boldsymbol{E} = -\mu_0 \frac{\partial^2 \Phi}{\partial z \partial t}\phi^{\,o} + \mu_0 \frac{1}{r}\frac{\partial^2 \Phi}{\partial \phi \partial t}\boldsymbol{z}^{\,o}\tag{52}$$

and

$$\boldsymbol{H} = -\left(\frac{1}{r^2}\frac{\partial^2 \Phi}{\partial \phi^2} + \frac{\partial^2 \Phi}{\partial z^2}\right)\boldsymbol{r}^{\,o} + \frac{1}{r}\left(\frac{\partial^2 \Phi}{\partial \phi \partial r} - \frac{1}{r}\frac{\partial \Phi}{\partial \phi}\right)\phi^{\,o} + \left(\frac{1}{r}\frac{\partial \Phi}{\partial z} + \frac{\partial^2 \Phi}{\partial r \partial z}\right)\boldsymbol{z}^{\,o},\tag{53}$$

respectively. If $\Phi$ are X wave solutions (eq.(20)), equations (52) and (53) represent limited diffraction electromagnetic X waves with their electric components perpendicular to $\boldsymbol{r}^{\,o}$.

For lower-order $\Phi_{XBB_n}$ (smaller $n$), $\boldsymbol{E}$ or $\boldsymbol{H}$ in eqs.(49)-(50) or eqs.(52)-(53) may be singular around the axial axis because of the terms $1/r$ and $1/r^2$.

## D. - Limited Diffraction and Wave Speed

Because $\boldsymbol{E}$ or $\boldsymbol{H}$ in eqs.(13)-(14), (49)-(50), or (52)-(53) are obtained by derivatives of the scalar X waves in terms of their spatial and time variables, the propagation term, $z - c_1 t$, is retained after the derivatives. Therefore, the electromagnetic X-shaped waves are also limited diffraction beams (traveling with the wave at speed $c_1$, one will see a constant wave pattern in space).

The wave speed, $c_1$, of limited diffraction beams (eq.(1)) along the $z$ axis is greater than or equal to the speed of sound (in acoustics) or the speed of light (in electromagnetism or optics). For example, the speed of the X-shaped waves is[36] $c_1 = c/\cos\zeta \geq c$, where $0 \leq \zeta < \pi/2$ is an Axicon angle[52]. This behavior is said to be "tachyonic" or Superluminal[53] and has been studied by many investigators in both "particle"[54–57] and wave [54,55,57,40–42,36] physics. In particular, recently (see e.g. refs.[40,41]) it has been claimed that all relativistic (homogeneous) wave equations admit also sub- and Super-luminal solutions: a claim later confirmed, e.g., in [57] (and that in the past had been verified only in some particular cases[54]).

Although the theoretical Superluminal waves encountered in this paper cannot be exactly produced, due to their infinite energy, they can be approximated with a finite aperture radiator and retain the essential characteristics (limited diffraction and Superluminal group velocity) over a large depth of field[36]. Each component (wavelet) of the approximated wave propagates at the speed of sound or of light, but the cone, or X-shaped wave, created by their superposition travels at a speed greater than the sound or light speed. The X waves, however, do not appear to violate[36,37,54,55] the special theory of relativity, as we shall discuss below in Section VII, when we shall also show that Superluminal X-shaped waves, in particular, are predicted by Relativity itself[53,54] to travel in space rigidly, without deforming.

Since Superluminal motions seem to appear even in other sectors of experimental science, we deem it proper to present in an Appendix at the end of this paper some information about those experimental results. In fact, the subject of Superluminal objects [or tachyons] addressed in this paper is still unconventional, and it can get more support from experiments than from theory. Moreover, such pieces of information are presently scattered in four different areas of science, and it can be useful to find them all collected in one and the same place.

*E. - Electric and Magnetic Field Strength Vectors*

For the electromagnetic X waves given by eqs.(22)-(26), $(\boldsymbol{E}_{X\mathrm{BB}_n})_{\phi}/(\boldsymbol{E}_{X\mathrm{BB}_n})_r$ or $(\boldsymbol{H}_{X\mathrm{BB}_n})_{\phi}/(\boldsymbol{H}_{X\mathrm{BB}_n})_r$ is not a function of $r$ and $\phi$. In addition, the $z$ component of $\boldsymbol{H}$ is very small as compared to the other two components (see Table I). This means that the major component of the Poynting flux (eqs.(27)-(29)) is in the $z$ direction (the direction of the vector Hertz potential) (see Table II).

*F. - Method to Obtain Other Limited Diffraction Electromagnetic Waves*

It is clear from equations (13)-(14), (49)-(50), or (52)-(53), that when new limited diffraction solutions to eq.(15) are found, the corresponding limited diffraction electromagnetic waves can be obtained. There are many ways to obtain new limited diffraction solutions to the scalar wave equation, such as the variable substitution method, that converts any existing solutions to a limited diffraction solution[24]; and the superposition method, that uses Bessel beams or rather X waves as basis functions to construct limited diffraction beams of practical usefulness[6,26].

*G. - Bowtie X Waves for Sidelobe Reduction*

Sidelobes of the scalar X-shaped waves (eq.(19)) are high along the X branches[4]. The asymptotic behavior of the electromagnetic X wave sidelobes are similar to that of the scalar waves (see Figs.1 and 2). Low sidelobes are necessary to obtain high contrast medical imaging where the biological soft tissues are modeled as multiple random scatterers[51]. To reduce sidelobes of the scalar X waves in pulse-echo imaging, a summation-subtraction method[31] can be used. However, this method reduces the frame rate in real-time pulse-echo imaging, and is also sensitive to object motion. Recently, unsymmetrical limited diffraction beams such as bowtie X waves were developed[4]. These waves are obtained by taking derivatives in one direction, say $y$, of the zeroth-order X wave. When a bowtie X wave is used to transmit and its 90° rotated beam pattern (around the $z$ axis) is used to receive, sidelobes of pulse-echo imaging systems can be reduced dramatically without compromising the image frame rate. The same technique could also be applied to the electromagnetic X-shaped waves for low sidelobe imaging.

## H. - Total Energy, Energy Density, and Causality

Because of the slow decay of the field along the X branches, the total energy (eq.(31)) of the abovementioned electromagnetic X-shaped waves is infinite. In addition, such waves should already exist at $t = -\infty$, due to the infinite extension of the X arms. Therefore, theoretical (infinitely extended) electromagnetic X waves are not experimentally realizable. However, because the energy density of the electromagnetic X waves is everywhere finite (except for those discussed in subsection C of this Section), these waves can be approximated very well over a large depth of interest (depth of field) by truncating them in both aperture (transverse plane) and axial range[36,37,58]. The truncated waves are "causal", in the sense that they do not meet the above problems.

## I. - Possible Applications

Because limited diffraction electromagnetic waves can be approximated over a large depth of field with a zero diffraction angle, they could have applications in acoustics[4–17], electromagnetism[18–20], and optics[2,3,21].

### *J. - Particle Like Character*

Limited diffraction X waves (eq.(19) and (41)) are particle like. As the scaling parameter, $a_0$, decreases, the wave-function energy density around the wave center increases. The off-center energy density of the X-shaped waves decays slowly along the X branches, which may provide a way to communicate with other wave particles[58]. Let us also notice that the X-shaped waves considered by us are not normalizable, like the plane wave solution[48] to the free Schroedinger equation.

### VII. — A THEORETICAL FRAMEWORK WITHIN SPECIAL RELATIVITY FOR THE "X-WAVES"

Let us here mention that a simple theoretical framework exists[54] (merely based on the space-time geometrical methods of Special Relativity (SR)) which incorporates the Superluminal X-shaped waves without violating the Relativity principles.

Actually, SR can be derived from postulating: (i) the Principle of relativity; and (ii) space-time to be homogeneous and space isotropic. It follows that one and only one *invariant* speed exists; and experience shows that invariant speed to be the one, *c,* of light in vacuum (the essential role of *c* in SR is just due to its invariance, and not to its being supposedly a maximal, or minimal, speed; no other, sub- or Super-luminal, object can be endowed with an invariant speed: in other words, no bradyon or tachyon can play in SR the same essential role as the speed-*c* light waves). Let us recall, incidentally, that tachyon [a term coined in 1967 by G.Feinberg] and bradyon [term coined in 1970 by one of the present authors] mean Superluminal and subluminal object, respectively. The speed $c$ turns out to be a limiting speed: but any limit possesses two sides, and can be approached a priori both from below and from above (as E.C.G.Sudarshan put it, from the fact that no one could climb over the Himalayas ranges, people of India cannot conclude that there are no people North of the Himalayas...; actually, speed-$c$ photons exist, which are born, live and die just "at the top of the mountain," without any need to accelerate from rest to the light speed).

A consequence is that the quadratic form $ds^2 = c^2 dt^2 - dx^2 - dy^2 - dz^2$ (i.e., the four-dimensional length-element square, along the space-time path of any object) results to be invariant, except for its sign. In correspondence with the positive (negative) sign, one gets the subluminal (Superluminal) Lorentz transformations [LT]. The ordinary, subluminal LTs leave, e.g., the fourvector squares and the scalar products (between fourvectors) just invariant.

The Superluminal LTs can be easily written down only in two dimensions (or in six, or in eight, dimensions...). But they must have the properties of changing sign, e.g., to the fourvector squares and to the fourvector scalar products. This is enough to deduce —see Fig.4— that a particle, which is spherical when at rest (and which appears then as ellipsoidal, due to Lorentz contraction, at subluminal speeds $v$), will appear[53,54,58,59] as occupying the cylindrically symmetrical region bounded by a two-sheeted rotation hyperboloid and an indefinite double cone, as in Fig.4(d), for Superluminal speeds $V$. In Fig.4 the motion is along the $x$-axis. In the limiting case of a point-like particle, one obtains only a double cone. In 1982, therefore, it was predicted[53] that the simplest Superluminal object appears (not as a particle, but as a field or rather) as a wave:

namely, as a "X-shaped wave", the cone semi-angle $\alpha$ being given (with $c = 1$) by $\cot \alpha = \sqrt{V^2 - 1}$.

It was also predicted[53,59] that the X-shaped waves would move *rigidly* with speed *V* along their motion direction (Fig.5). The reason for such "X-waves" to travel undeformed is quite simple: every X-wave can be regarded at each instant of time as the (Superluminal) Lorentz transform of a spherical object, which of course as time elapses moves in vacuum without any deformation.

The X-shaped waves here considered are the most simple ones only. If we started not from an intrinsically spherical or point-like object, but from a non-spherically symmetric particle, or from a pulsating (contracting and dilating) sphere, or from a particle oscillating back and forth along the motion direction, their Superluminal Lorentz transforms would result to be more and more complicated. The above-seen X-waves, however, are typical for a Superluminal object, so as the spherical or point-like shape is typical for a subluminal particle.

The three dimensional picture of Fig.5, or rather of Fig.4(d), appears in Fig.6, where its annular intersections with a transverse plane are shown (cf. refs.[53,59]).

It has been believed for a long time that Superluminal objects would have allowed sending information into the past; but such problems with causality seem to be solvable within SR. Once SR is generalized in order to include tachyons, no signal traveling backwards in time is apparently left. For a solution of those causal paradoxes, see refs.[55,56] and references therein.

Let us pass, within this elementary context, to the problem of producing a "X-shaped wave" like the one depicted in Fig.6 (truncated of course, in space and in time, by use of a finite antenna radiating for a finite time interval). To convince ourselves about the possibility of realizing them, it is enough to consider *naively* the ideal case of a Superluminal source $S$ of negligible size, endowed with constant speed $V$ and emitting spherical electromagnetic waves $W$ (each one traveling at the invariant speed $c$). We shall observe the electromagnetic waves to be internally tangent to an enveloping cone $C$ having the source motion line $x$ as its axis and $S$ as its vertex. This is analogous to what happens with an airplane moving with a constant supersonic speed in the air. In addition, those electromagnetic waves $W$ interfere negatively one another inside the cone $C$, and interfere constructively only on its surface. We can put a plane detector orthogonal to $x$ and record the intensity of the waves $W$ impinging on it, as a (cylindrically symmetric) function of position and of time. Afterwards, it will be enough to replace the plane detector by a plane antenna that radiates —instead of detecting and recording— exactly the same (cylindrically symmetrical) space-time pattern of electromagnetic waves $W$, in order to build up a cone-shaped ($C$) electromagnetic wave travelling along $x$ with the Superluminal speed $V$ (obviously, with no radiating source —now— any longer at its vertex $S$). Even if each spherical wave $W$ will still travel with the invariant speed $c$.

Incidentally, by evaluation of the above-mentioned intensity, one can get the simplest example of "X-waxe solution."

Let us recall by the way that, in the approximated case in which we produce a finite conic wave truncated both in space and in time, the theory of SR suggested the bi-conic shape (symmetrical in space with respect to the vertex $S$) to be a priori a better approximation to a rigidly traveling wave; so that SR suggests to have recourse to an antenna emitting a radiation (not only cylindrically symmetrical in space but also) symmetric in time, in order to obtain *a priori* a better approximation to an undeformed progressive wave.

Let us mention moreover that our finite bi-conic, or X-shaped, waves (after having been produced) are expected to travel almost rigidly, at Superluminal speed, without any further support from the radiator.

One may observe, at last, that in the vacuum and in nondispersive media for our X-shaped waves the group velocity coincides with the phase velocity.

## VIII. — CONCLUSION

We have derived limited diffraction solutions to the free Maxwell equations. Theoretically, these solutions are diffraction-free. They are infinitely extended in space and time, and possess infinite total energy. However, for solutions that are single valued and non-singular (finite energy density), they can be approximated very well by a finite aperture antenna over a large depth of interest. In addition, some X-shaped wave solutions to the free Schroedinger equation are derived; which may be helpful for a better understanding of the relationship between waves and particles. Because limited diffraction beams have a large depth of field, they could have applications in several areas.

As explained in Sect.VI-D above, we use the present occasion to present in the Appendix below some information about the other sectors of experimental science in which Superluminal motions seem to appear.

## IX. — ACKNOWLEDGMENTS

The authors appreciate for helpful discussions Dr. Hugo E. Hernandez, Dr. L. C. Kretly and Dr. Waldyr A. Rodrigues Jr. of the D.M.O.-FEE, of the C.C.S. and of IMECC, respectively, all at the Universidade Estadual de Campinas, Campinas, S.P., Brazil. This work was supported in part by grants CA54212 and CA43920 from the National Institutes of Health (as well as by CNPq, and by CNR, MURST, INFN).

## X. — APPENDIX

In this Appendix we take the opportunity to present some sketchy information —mainly bibliographical— about the other three (in total, four) sectors of experimental science in which Superluminal motions seem to appear. In fact, as the "Superluminal" topic is still controversial, a panoramic view of the overall *experimental* situation is certainly useful, especially when it is considered that the related information, scattered in very different Journals, is not all of easy access to everybody.

For the sake of brevity, in the references quoted in this Appendix the title of the papers will be omitted.

The question of Superluminal objects or waves [tachyons] has a long story, starting perhaps with Lucretius' *De Rerum Natura.* Still in pre-relativistic times, let us recall e.g. the contributions by A.Sommerfeld. In relativistic times, our problem started to be tackled again essentially in the fifties and sixties, in particular after the papers by E.C.George Sudarshan et al., and later on by E.Recami, R.Mignani et al. [who by their numerous works at the beginning of the seventies rendered, by the way, the terms sub- and Super-luminal of popular use], as well as by H.C.Corben and others (to confine ourselves to the theoretical researches). For references, one can check pages 162-178 in ref.[54], where about 600 citations are listed; or the large bibliographies by V.F.Perepelitsa[60] and the book in ref.[61]. In particular, for the causality problems one can see refs.[55,56] and references therein, while for a model theory for tachyons in two dimensions one can be addressed to refs.[54,62]. The first experiments looking for tachyons were performed by T.Alvager et al.; some citations about the early experimental quest for Superluminal objects being found e.g. in refs.[1,63].

The subject of tachyons is presently returning after fashion, especially because of the fact that four different experimental sectors of physics seem to indicate the existence of Superluminal objects: including —of course— the one dealt with in this and in other papers of ours, which seems to us as being *at the moment* the most important sector. Let us put forth in the following some brief information about the experimental results obtained in such different science areas.

*First: Negative Mass-Square Neutrinos*

Since 1971 it was known that the experimental square-mass of muon-neutrinos resulted to be negative[64]. If confirmed, this would correspond (within the ordinary naive approach to relativistic particles) to an imaginary mass and therefore to a Superluminal speed; in a non-naive approach[54], i.e. within a Special Relativity theory extended to include tachyons (Extended Relativity), the free tachyon "dispersion relation" (with $c = 1$) becomes $E^2 - \boldsymbol{p}^2 = -m_0^2$.

From the theoretical point of view, let us refer to [65] and references therein.

Recent experiments showed that also electron-neutrinos result to have negative mass-square[66].

*Second:  Galactic "Mini-Quasars"*

We refer ourselves to the apparent Superluminal expansions observed inside quasars, some galaxies, and —as discovered very recently— in some galactic objects, preliminarily called "mini-quasars". Since 1971 in many quasars (and even a few galaxies) apparent Superluminal expansions were observed [*Nature,* for instance, dedicated to those observations a couple of its covers]. Such seemingly Superluminal expansions were the consequence of the experimentally measured angular separation rates, once it was taken into account the (large) distance of the sources from the Earth. From the experimental point of view, it will be enough to quote the book [67] and references therein.

The distance of those "Superluminal sources", however, it is not well known; or, at least, the (large) distances usually adopted have been strongly criticized by H.Arp et al., who maintain that quasars are much nearer objects than expected: so that all the above-mentioned data can no longer be easily used to infer (apparent) Superluminal motions. However, very recently, *galactic* objects have been discovered, in which apparent Superluminal expansions occur; and the distance of galactic objects can be more precisely determined. From the experimental point of view, see in fact the papers [68].

From the theoretical point of view, both for quasars and "mini-quasars", see [69,54]. In particular, let us recall that a *single* Superluminal source of light would be observed: (i) initially, in the phase of "optic boom" (analogous to the acoustic "boom" by an aircraft that travels with constant super-sonic speed) as a suddenly-appearing, intense source; (ii) later on, as a source which splits into TWO objects receding one from the other with relative velocity *V* larger than *2c.*

*Third:  Tunnelling photons = Evanescent waves*

It is the sector that most attracted the attention of the scientific and non-scientific *press*[70].

Evanescent waves were predicted [cf., e.g., ref.[54], page 158 and references therein] to be faster-than-light.  Even more, they consist in tunnelling photons:  and it was known since long time[71] that tunnelling particles (wave packets) can move with Superluminal group velocities inside the barrier; therefore, due to the theoretical analogies between tunnelling particles (e.g., electrons) and tunnelling photons[49], it was since long expected that evanescent waves could be Superluminal.

The first experiments have been performed at Cologne, Germany, by Guenter Nimtz et al., and published in 1992.  See refs.[72].

Other very famous experiments have been performed at Berkeley: see refs.[73].

Further experiments on Superluminal evanescent waves have been done at Florence[74]; while a last experiment (as far as we know) took place at Vienna[74].

From the theoretical point of view, see [71] and references therein; and [75].

*Fourth:  Superluminal motions in Electrical and Acoustical Engineering — The "X-shaped waves"*

This fourth sector, which this paper is contributing to, seems to be at the moment

(together with the third one) the most promising.

Starting with the pioneering work by H.Bateman, it became slowly known that all the relativistic homogeneous wave equations —in a general sense: scalar, electromagnetic and spinor— admit solutions with subluminal group velocities[76]. More recently, also Superluminal solutions have been constructed for those homogeneous wave equations, in refs.[77,57] and quite independently in refs.[40-42]: in some cases just by applying a Superluminal Lorentz "transformation"[54,62]. It has been also shown that an analogous situation is met even for acoustic waves, with the existence in this case of "sub-sonic" and "Super-sonic" solutions[10,36]; so that one can expect they to exist, e.g., also for seismic wave equations. More intriguingly, we may expect the same to be true in the case of gravitational waves too.

Let us recall that the rigidly traveling Supersonic and Superluminal solutions found in refs.[36,37] and in this paper —some of them already experimentally realized— appear to be (generally speaking) X-shaped, just as predicted in 1982 by Barut, Maccarrone and Recami[53].

On this regard, from the theoretical point of view, let us quote pages 116-117, and pages 59 (fig.19) and 141 (fig.42), of ref.[54]; and even more refs.[54,58,59], where "X-shaped waves" are predicted and discussed. From such papers it is also clear why the X-shaped waves keeps their form while traveling (nondiffracting waves).

faster than light?" by R.Landauer, Oct. 21, 1993; *New Scientist:* editorial "Faster than Einstein" at p.3, plus an article by J.Brown at p.26, April 1995.

**FIGURE CAPTIONS**

**Fig.1.** Real part of Hertz potential and field components of the zeroth-order ($n = 0$) limited diffraction electromagnetic X wave at time $t = z/c_1$. Panel (1) is Hertz potential, $Re\{\Phi_{XBB_0}\}$ ; Panel (2) is the $\phi$ component of electric field strength, $Re\left\{(\boldsymbol{E}_{XBB_0})_\phi\right\}$ ; and Panels (3) and (4) are $r$ and $z$ components of magnetic field strength, $Re\{(\boldsymbol{H}_{XBB_0})_r\}$ and $Re\{(\boldsymbol{H}_{XBB_0})_z\}$, respectively. The dimension of each panel is 4 m ($r$ direction) $\times$ 2 mm ($z$ direction). The free parameters $\zeta$ and $a_0$ are 0.005° and 0.05 mm, respectively. The values shown on the right-top corner of each panel represent the maxima and the minima of the images before normalization for display [MKSA units] (see also Table I).

**Fig.2.** Poynting flux and energy density of the zeroth-order limited diffraction electromagnetic X wave at time $t = z/c_1$. Panels (1) and (2) are $r$ and $z$ components of the Poynting flux, $(\boldsymbol{P}_{XBB_0})_r$ and $(\boldsymbol{P}_{XBB_0})_z$, respectively; and Panel (3) is the energy density, $U_{XBB_0}$. The dimension of each panel and the parameters of the X waves are the same as those in Fig.1. The values shown on the right-top corner of each panel represent the maxima and the minima of the images before normalizing for display [MKSA units] (see also Table II).

**Fig.3.** Line plots of the zeroth-order electromagnetic X wave in Figs.1 and 2 along one of the "X" branches (from left bottom to top right). Panel (1) shows the line plots of the field components: $Re\{\Phi_{XBB_0}\}$ (full line), $Re\left\{(\boldsymbol{E}_{XBB_0})_\phi\right\}$ (dotted line), $Re\{(\boldsymbol{H}_{XBB_0})_r\}$ (dashed line), and $Re\{(\boldsymbol{H}_{XBB_0})_z\}$ (long dashed line). Panel (2) is the line plot of the Poynting flux and energy density: $(\boldsymbol{P}_{X\mathrm{BB}_0})_r$ (full line), $(\boldsymbol{P}_{X\mathrm{BB}_0})_z$ (dotted lines), and $U_{X\mathrm{BB}_0}$ (dashed line).

**Fig.4.** Let us consider an object that is intrinsically spherical, i.e., that is a sphere in its rest-frame (Panel (a)). After a generic subluminal Lorentz transformation (LT) along $x$, i.e., under a subluminal $x$-boost, it is predicted by special relativity (SR) to appear as ellipsoidal due to Lorentz contraction (Panel (b)). After a Superluminal $x$-boost (namely, when this object moves with Superluminal speed $V$), it is predicted by extended relativity (ER) to appear[53,54] as in Panel (d), i.e., as occupying the cylindrically symmetric region bounded by a two-sheeted rotation hyperboloid and an indefinite double cone. The whole structure is predicted by ER to move rigidly and, of course, with the speed $V$, the cone semi-angle cotangent square being $(V/c)^2 - 1$. Panel (c) refers to the limiting case when the boost-speed tends to $c$, either from the left or from the right. (For simplicity, a space axis is skipped).

**Fig.5.** If we start from a spherical particle as in Fig.4(a), then —after a Superluminal boost along a generic motion line $l$,— we obtain the tachyonic object $T$ depicted in this figure. Once more, the Superluminal object $T$ appears to be spread over the whole spatial region delimited by a double cone and a two-sheeted hyperboloid asymptotic to the cone[59]. The whole structure travels of course along $l$ with the speed $V$ of the Superluminal $l$-boost. Notice that, if the object is not spherical when at rest (but, e.g., is ellipsoidal in its own rest-frame), then the axis of $T$ will no longer coincide with $l$, but its direction will depend on the speed $V$ of the tachyon itself. For the case in which the space extension of the Superluminal object $T$ is finite, see ref.[58].

**Fig.6.** Here we show the intersections of the Superluminal object $T$ with planes $P$ orthogonal to its motion line $x$-axis, for the same case as in Fig.4. For simplicity, we assumed again the object to be spherical in its rest-frame, and the cone vertex to coincide with the origin O for $t = 0$. Such intersections evolve in time so that the same pattern appears on a second plane —shifted by $\triangle x$ after the time $\triangle t = \triangle x/V$. On each plane, as time elapses, the intersection is therefore predicted by ER to be a circular ring which, for negative times, goes on shrinking until it reduces to a circle and then to a point (for $t = 0$); afterwards, such a point becomes again a circle and then a circular ring that goes on broadening[53,54,59,58].

**TABLE CAPTIONS**

**Table I:** Maxima and minima of the zeroth-order limited diffraction electromagnetic X waves (unit: MKSA).

|      | $Re\{\Phi_{XBB_0}\}$ | $Re\{(\boldsymbol{E}_{XBB_0})_\phi\}$ | $Re\{(\boldsymbol{H}_{XBB_0})_r\}$ | $Re\{(\boldsymbol{H}_{XBB_0})_z\}$ |
|------|------|------|------|------|
| max  | 1.0  | $9.5\times10^6$ | $2.5\times10^4$ | 6.1 |
| min  | 0.0  | $-9.5\times10^6$ | $-2.5\times10^4$ | -1.5 |

**Table II:** Maxima and minima of the Poynting flux and energy density of the zeroth-order limited diffraction electromagnetic X waves (unit: MKSA).

|      | $(\boldsymbol{P}_{XBB_0})_r$ | $(\boldsymbol{P}_{XBB_0})_z$ | $U_{XBB_0}$ |
|------|------|------|------|
| max  | $2.4\times10^7$ | $2.4\times10^{11}$ | $1.6\times10^3$ |
| min  | $-2.4\times10^7$ | 0.0 | 0.0 |

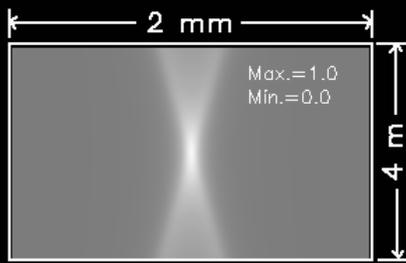
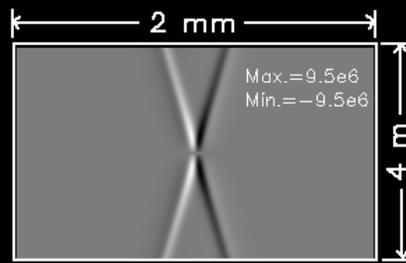
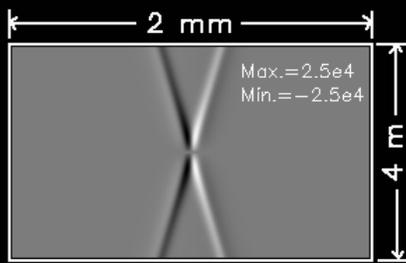
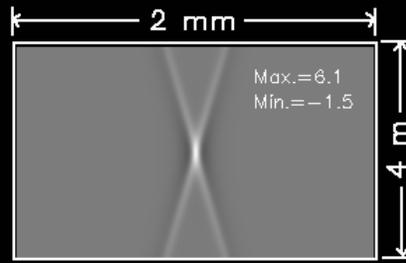
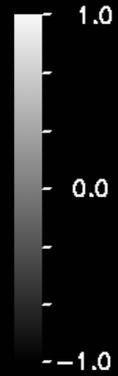

(1) $R_e\{\underline{\Phi}_{XBB_0}\}$

(2) $R_e\{(\vec{E}_{XBB_0})_\theta\}$

(3) $R_e\{(\vec{B}_{XBB_0})_\rho\}$

(4) $R_e\{(\vec{B}_{XBB_0})_z\}$

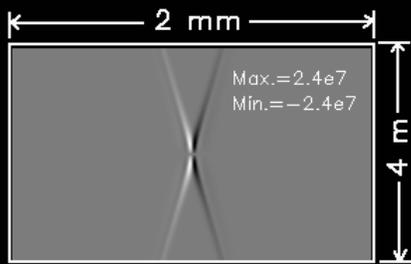
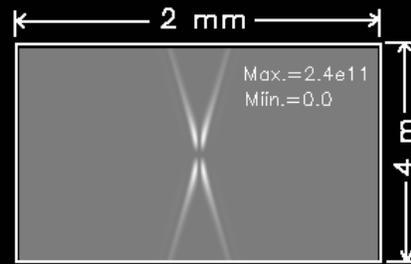
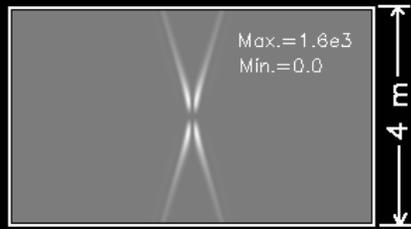
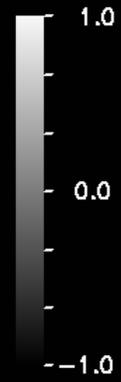

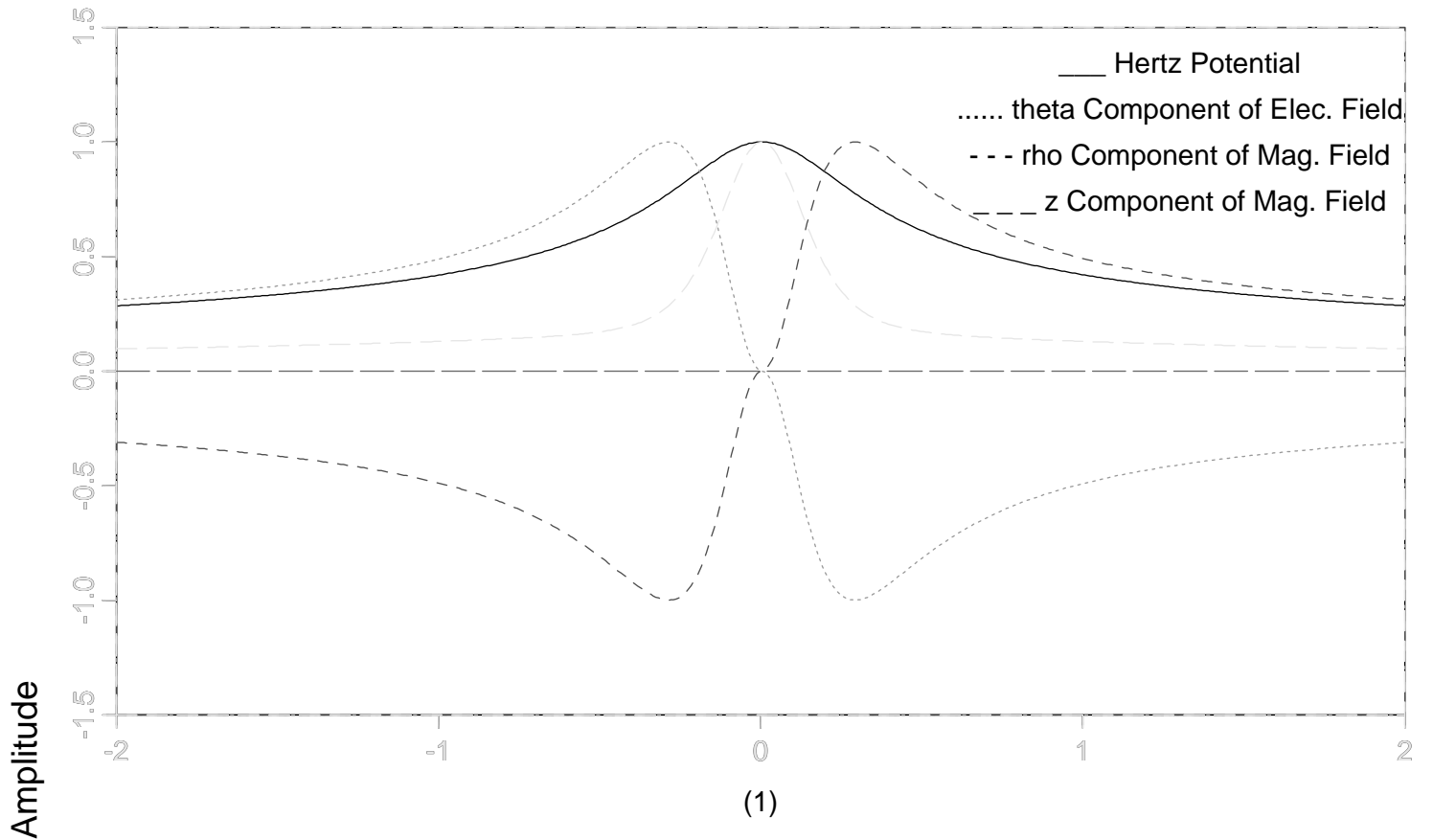

(1)

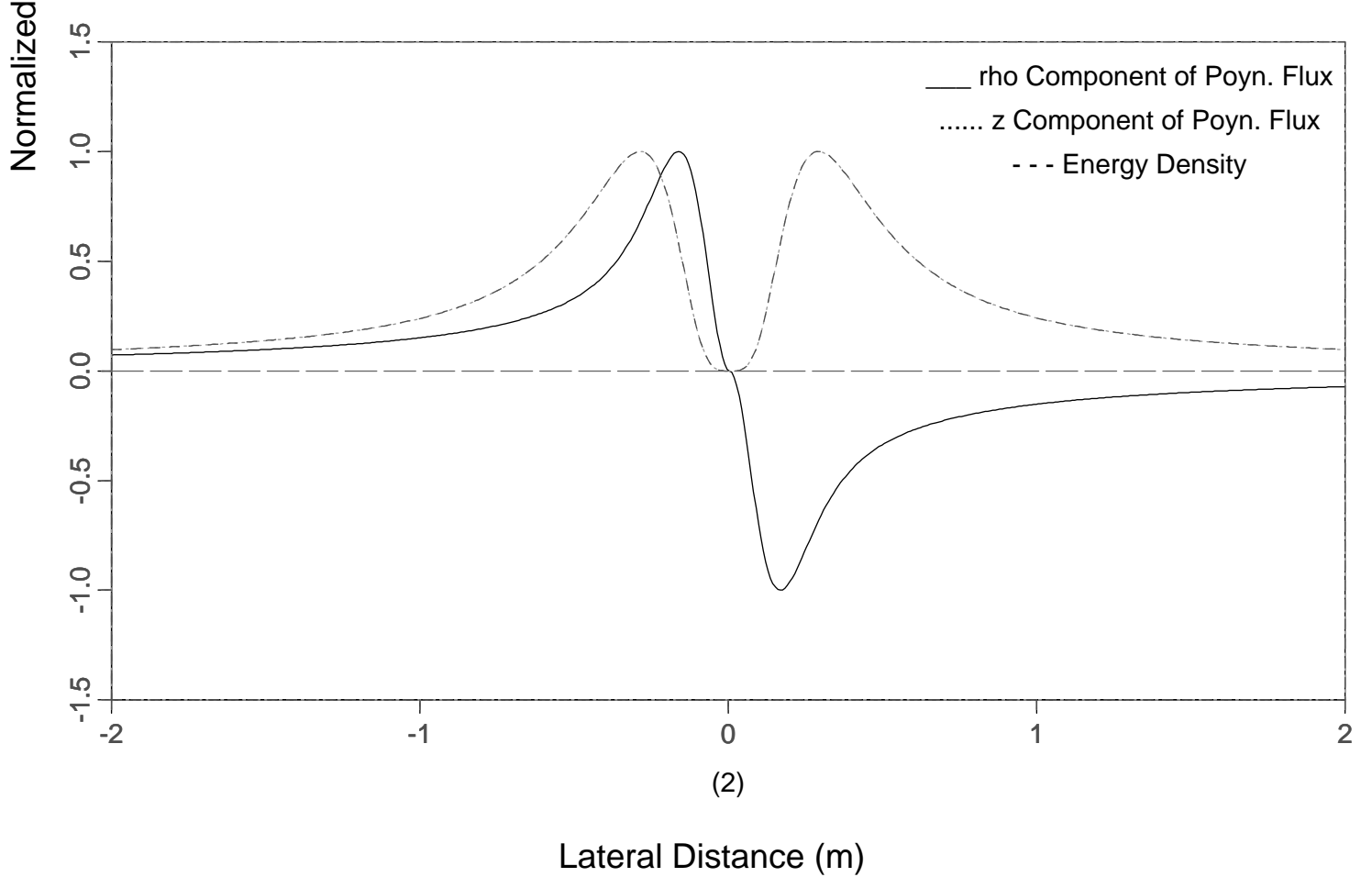

(2)

Lateral Distance (m)

Normalized Amplitude